# Structural and optical properties of self-assembled AlN nanowires grown on SiO$_2$/Si substrates by molecular beam epitaxy

Ž. Gačević,[1] J. Grandal,[1,2] Q. Guo,[3] R. Kirste,[3] M. Varela,[2] Z. Sitar[3], M.A. Sánchez García[1]

[1] ISOM, Universidad Politécnica de Madrid Avda. Complutense 30, 28040 Madrid, Spain
[2] GFMC, Departamento de Física de los Materiales & Instituto Pluridisciplinar, Universidad Complutense de Madrid 28040 Spain
[3] Department of Materials Science and Engineering, North Carolina State University, Raleigh, NC 27695-7919, USA

**Abstract**

Self-assembled AlN nanowires (NWs) are grown by plasma-assisted molecular beam epitaxy (PAMBE) on SiO$_2$ / Si (111) substrates. Using a combination of *in-situ* reflective high energy electron diffraction and *ex situ* X-ray diffraction (XRD), we show that the NWs grow nearly strain-free, preferentially perpendicular to the amorphous SiO$_2$ interlayer and without epitaxial relationship to Si(111) substrate, as expected. Scanning electron microscopy investigation reveals significant NWs coalescence, which results in their progressively increasing diameter and formation of columnar structures with non-hexagonal cross-section. Making use of scanning transmission electron microscopy (STEM), the NWs initial diameters are found in the 20 – 30 nm range. In addition, the formation of a thin (≈30 nm) polycrystalline AlN layer is observed on the substrate surface. Regarding the structural quality of the AlN NWs, STEM measurements reveal the formation of extended columnar regions, which grow with a virtually perfect metal-polarity wurtzite arrangement and with extended defects only sporadically observed. Combination of STEM and electron energy loss spectroscopy (EELS) reveals the formation of continuous aluminum oxide (1–2 nm) on the NW surface. Low temperature photoluminescence measurements reveal a single near-band-edge (NBE) emission peak, positioned at 6.03 eV (at 2 K), a value consistent with nearly zero NW strain evidenced by XRD and in agreement with the values obtained on AlN bulk layers synthesized by other growth techniques. The significant full-width-at-half-maximum of NBE emission, found at ≈20 meV (at 2 K), suggests that free and bound excitons are mixed together within this single emission band. Finally, the optical properties of the hereby reported AlN nanowires grown by PAMBE are comprehensively compared to optical properties of bulk, epitaxial and/or columnar AlN grown by various techniques such as: physical vapor transport (PVT), metal organic vapor phase epitaxy (MOVPE), metal organic chemical vapor deposition (MOCVD) and molecular beam epitaxy (MBE).

The realization of highly efficient III-nitride light emitting and laser diodes has led to a revolution in the solid state lighting market [1]. Nowadays, III-nitrides are of major importance for blue and near UV light emitting devices [2]. Their ultimate potential, however, might be significantly beyond the limits of their current applications. One of the main constraints for their further development remains the lack of substrates with adapted crystal lattice, resulting in growth of defective III-nitride compact layers (dislocation density typically ≈$10^9$ cm$^{-2}$) [3]. As an alternative to defective III-nitride compact layers, in the late 1990s several groups reported self-assembled (SA) GaN nanowires (NWs), which grow with supreme crystal quality [4, 5]. These early results on SA NWs were later extended to the achievement of selective area grown GaN NWs, grown in highly uniform ensembles [6, 7].

While growth of GaN NWs has been thoroughly studied in the last two decades, the growth of AlN NWs is less known. The realization of high quality AlN NWs might facilitate development of optoelectronic devices further into UV range, i.e. beyond the GaN band gap limit (3.4 – 6.0 eV / 360 – 200 nm). Indeed, it has been shown that (Al,Ga)N NWs can be used for the realization of both highly efficient UV LEDs and low-threshold current UV lasers [8 - 10]. Another particularly interesting and promising applications of NWs is related to the realization of quantum light emitters operating at room temperature. Making use of high (Al,Ga)N/GaN band offsets, single photon source operation has already been demonstrated in the UV range and up to 350 K [11]. Substitution of (Al)GaN columnar barriers with their AlN counterparts would bring several additional benefits to SPS performance: it would allow single photon emission extension further into UV region, help for single photon emission purity and facilitate their operation at even higher temperatures.

The realization of SA GaN NWs on Si substrates by molecular beam epitaxy (MBE) relies on poor GaN wetting of the underlying surface. GaN, thus, tends to nucleate in the form of 3D islands, which via preferential growth in the vertical direction finally convert into columnar structures [5]. AlN, on the other hand, "wets" very well the underlying Si surface, extending into compact 2D layers, making thus realization of AlN columnar structure directly on Si a very difficult task. It has been shown that one possibility to significantly reduce AlN wetting (and consequently achieve columnar AlN) is to perform growth on thin silicon oxide layer instead [12]. In the present work, we adopt this approach to study the growth of SA AlN NWs and their structural and optical properties.

AlN NWs were grown in a plasma-assisted MBE system, on $SiO_2$/Si(111) substrates. To enable 3D AlN nucleation, prior to the MBE growth, a ≈500 nm thick amorphous $SiO_2$ layer was thermally deposited on commercial Si(111) substrate. To provide highly N-rich conditions for columnar growth, impinging Al and active N equivalent fluxes were set to ≈3 nm/min and ≈6 nm/min, respectively, the growth temperature was set to ≈ 920 - 940 °C whereas the growth time was set to 3 hours.

*In situ* reflection high energy electron diffraction (RHEED), performed with an electron beam accelerated to 15 keV, reveals that the AlN grows preferentially perpendicular to the $SiO_2$ surface and with hexagonal wurtzite crystal structure (Fig. 1(a)). The RHEED pattern, taken along $[11\bar{2}0]$ crystallographic direction, also reveals that Bragg spots are in the form of rings, which is a fingerprint of high dispersion in crystallographic tilt (in this case, NWs tilt). In addition, the RHEED pattern is invariant to sample rotation, which is a fingerprint of random *in plane* crystallographic orientation (in this case, random NWs twist), confirming lack of any epitaxial relationship to the underlying Si(111) substrate [12].

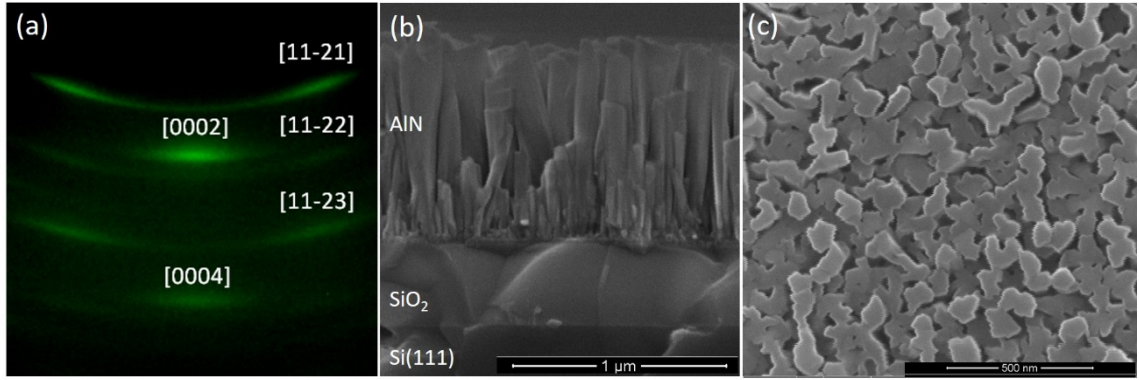

**Figure 1.** (a) RHEED image taken along $[11\bar{2}0]$ azimuth reveals that the NWs grow preferentially perpendicular to the $SiO_2$/Si(111) substrate surface but with high dispersion in NWs tilt. The invariance of the RHEED pattern to the sample rotation points to random NWs twist. (b) Cross-section SEM image reveals formation of columnar structures with progressively increasing diameter, whereas (c) top-view SEM image reveals formation of nanostructures, with non-hexagonal cross-section, both results being consequences of significant NWs coalescence.

Figures 1(b)-(c) show cross- and top-view scanning electron microcopy (SEM) images (performed by FEI Inspect F50 system) of one representative sample. The cross-view image reveals the formation of high density columnar structures with progressively increasing diameter. The increase in diameter is mainly attributed to the NWs coalescence, which becomes very significant at NWs heights of ≈300 nm. Bearing in mind that NWs nucleate with a random twist, the coalescence results in formation of AlN nanostructures with an irregular shape i.e. with non-hexagonal cross-section (commonly observed for the NWs constituted from crystals with six-fold wurtzite structure), as observed in top-view SEM images featured in Fig. 1(c).

To get a detailed insight into NWs structural and chemical properties, scanning transmission electron microscopy (STEM) and electron energy loss spectroscopy (EELS) were performed in an aberration-corrected JEOL ARM200cF electron microscope operated at 200kV equipped with a cold field emission gun, a spherical aberration corrector and a Gatan Quantum spectrometer. Samples were prepared by conventional mechanical grinding and Ar ion milling for the cross sectional measurements. In addition, some NWs were dispersed on a carbon grid in order to get a better understanding of the defects generated in individual NWs.

STEM examinations reveal the formation of relatively thin AlN NWs, with initial diameters in 20 - 30 nm range (Fig. 2(a)). Somewhat surprisingly, we find a thin (≈30 nm) AlN layer, formed on the SiO$_2$ surface, in between AlN NWs (a similar layer is not observed in a standard SA GaN NWs growth on Si) [13].

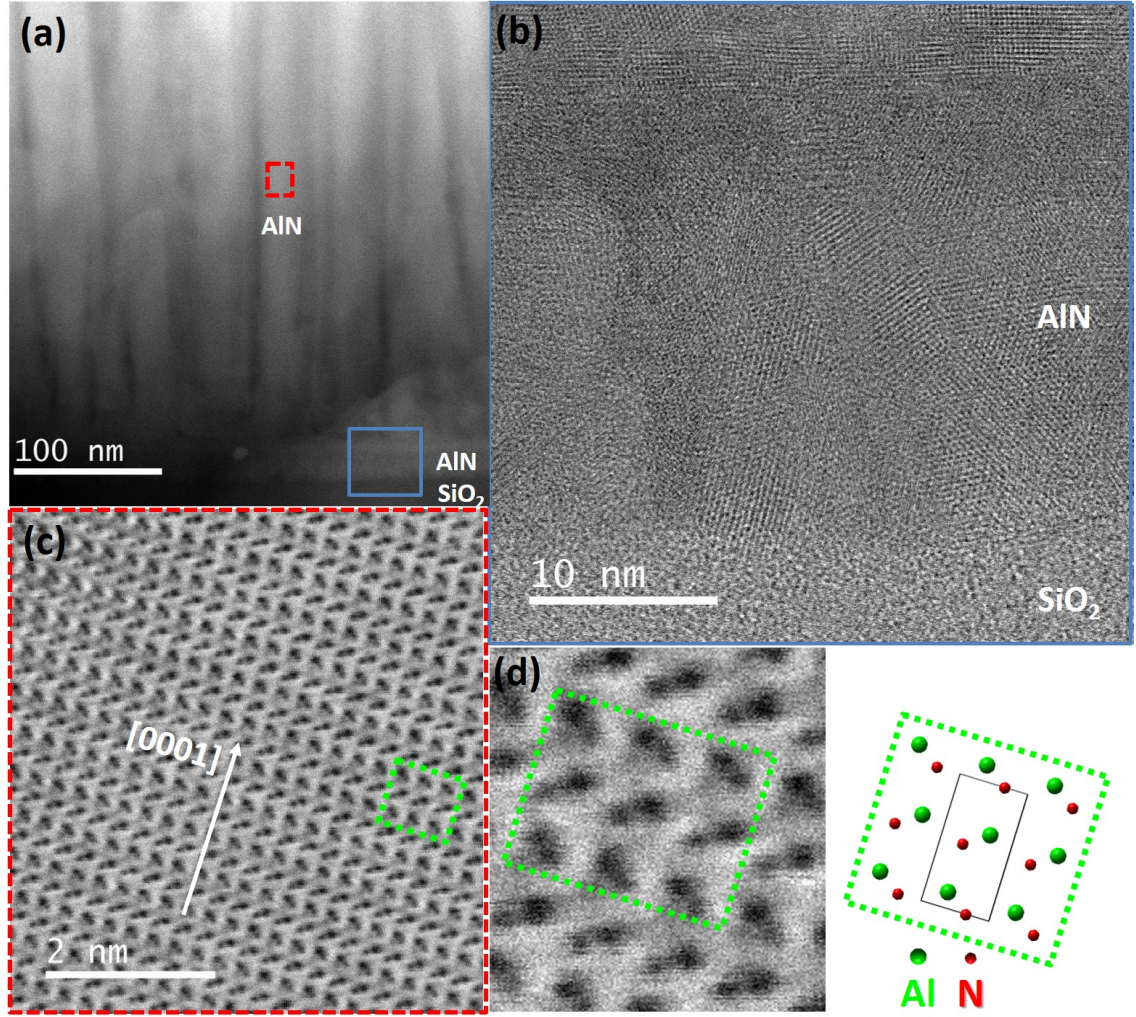

Figure 2. (a) STEM annular bright field (ABF) image of the AlN/SiO$_2$ interface reveals formation of a thin continuous AlN surface layer (designated with blue solid-line rectangle, for clarity). Nanowires initial diameters are estimated mainly within 20 – 30 nm range. (b) STEM ABF image of the AlN surface layer confirms amorphous nature of SiO$_2$ (the bottom part of the image) and polycrystalline nature of the AlN (the middle part of the image). (c) Atomic resolution ABF STEM image taken at a mid-height of a single representative nanowire (designated with red dashed-line square, for clarity) reveals extended regions of a perfect wurtzite crystalline structure. (d) A region designated with green dot-line rectangle, shown with higher magnification, reveals heavier (Al) atoms on top of lighter (N) atoms, within Al-N bilayers (a schematic is also shown for clarity). This configuration corresponds to Al-polarity.

Further high resolution examinations reveal the polycrystalline nature of this continuous layer (Fig. 2(b)). The formation of this layer, however, is not linked with the initial stages of NWs nucleation. It is important to note that the AlN NWs are grown at temperatures (920 – 940 °C) significantly below the AlN decomposition threshold (> 1000 °C) [14, 15]; this is not the case for SA GaN NWs, which are commonly grown above 800 °C, i.e. in the temperature range where GaN crystalline nuclei can be decomposed from the Si substrate and Ga atoms consequently desorbed from the surface [13, 14]. Thus, we link this thin layer to the "parasitic nucleation" of AlN on the $SiO_2$, that is, to the formation of AlN nuclei between already formed AlN NWs (in the later growth stages) and their consequent development into a continuous polycrystalline layer, as observed by STEM. The thickness of this layer, found at ≈30 nm, is most likely limited by the shadow effect imposed by the NWs, which are already in an advanced growth stage. STEM images with atomic resolution taken at NWs mid-height, reveal formation of extended regions with a perfect wurtzite structure (Fig. 2(c)). Close inspection of the image (Fig. 2(d)) reveals heavier atoms (Al) on the top of lighter atoms (N), within Al-N bilayers (see fig. 3(d) schematic, for clarity), proving that the NWs grow with Al polarity.

STEM measurements also reveal that despite the highly polycrystalline nature of the AlN epilayer grown on $SiO_2$, single NWs, emerging from the $SiO_2$ surface are found to be monocrystalline. Intra-wire extended defects are rarely observed. Figure 3(a) shows an atomic resolution annular bright field (ABF) STEM image, while 3(b) exhibits a fast Fourier filtered ABF image (removing [0002] planes) of selected AlN NW in the proximity of a single stacking fault (SF).

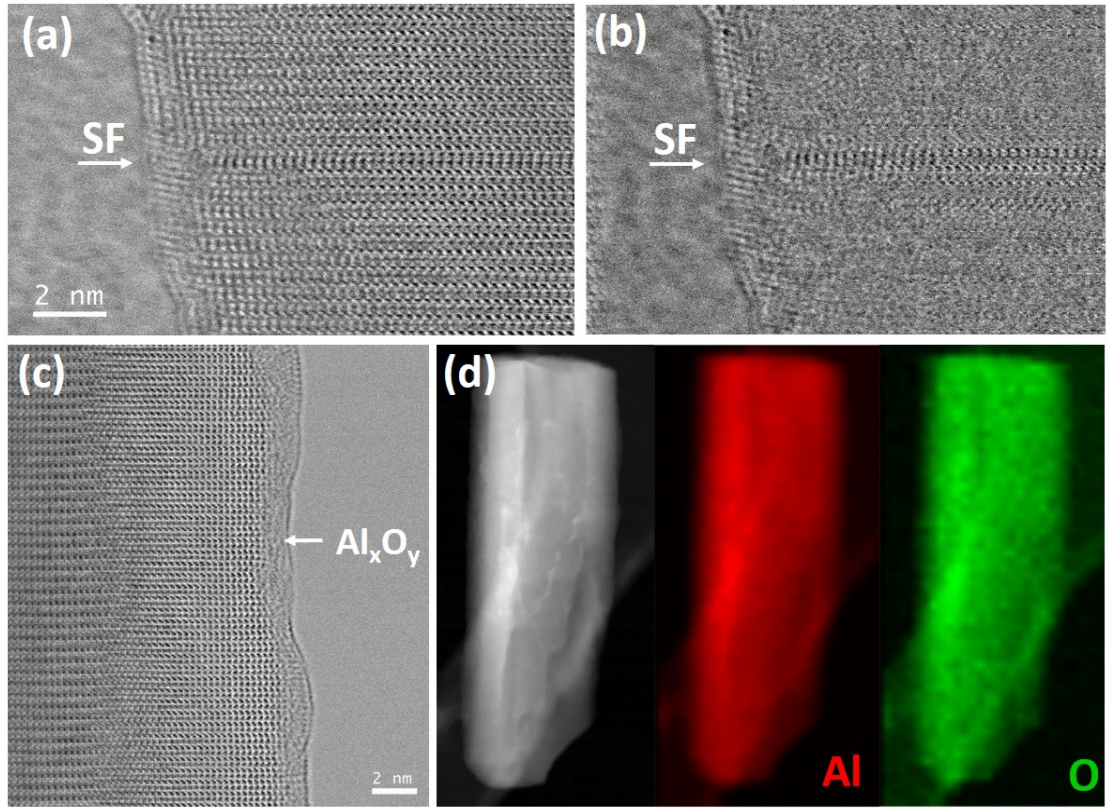

Figure 3. (a) Atomic resolution STEM image and (b) fast Fourier filtered STEM image removing [0002] planes revealing formation of a stacking fault. (c) Identification of 1-2 nm thick $Al_xO_y$ layer on the NW surface and (d) ADF and EELS elemental maps extracted from the Al $K$ and O $K$ edges of single lying AlN NW, further confirming $Al_xO_y$ formation.

Unlike GaN, AlN is prone to surface oxidation. Consequently, a thin (1 – 2 nm thick) $Al_xO_y$ wrapping layer has been observed around NWs (Fig. 3(c)) by STEM. To confirm the chemical composition of the observed wrapping layer with certainty, EELS measurements on single NWs have been performed. EEL spectra confirm the presence of Al, N and O atoms (results not shown). Fig. 3(d) shows the EELS chemical maps of the Al *K* (red) and O *K* (green) edges of an atomic resolution 2D EEL-spectrum image, along with the simultaneously acquired annular dark field (ADF) of a single NW. These maps exhibit a highly homogeneous spatial distribution of aluminum and oxygen along the whole NW, respectively.

To assess crystallographic properties of the as-grown NW ensembles, X-ray diffraction (XRD) measurements were performed in a commercial Panalytical X'Pert Pro diffractometer with Cu-$K_{\alpha 1}$ line ($\lambda_X$ = 1.5406 Å), provided by a parabolic mirror collimator followed by a channel-cut four-bounce Ge(220) monochromator, on the X-ray source side [16].

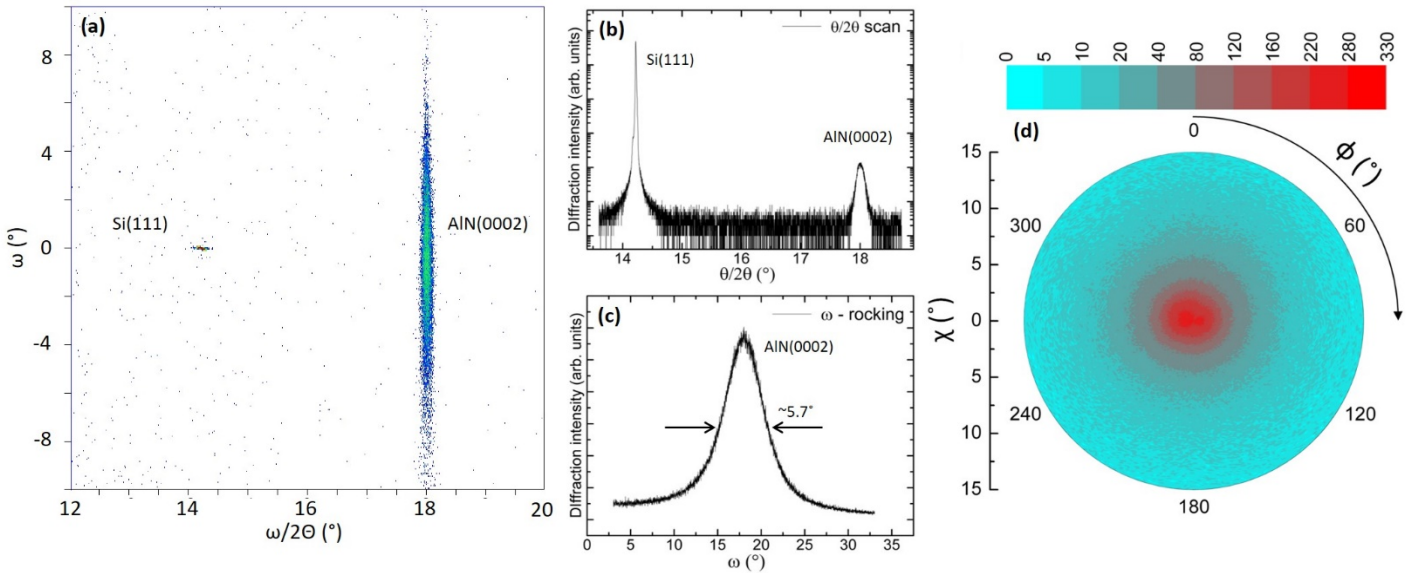

Figure 4. XRD study of a representative sample. (a) 2D ω/2Θ – ω scan performed in the proximity of the (111) Bragg reflection of the underlying Si(111) substrate reveals (0002) Bragg reflection of AlN and no Bragg reflections of $SiO_2$ (b) θ/2θ scan performed in the proximity of Si(111) Bragg reflection (i.e. along ω = 0° line in the 2D map) reveals that AlN NWs grow nearly fully relaxed. (c) ω-rocking scan around AlN(0002) Bragg reflection (i.e. along ω/2Θ = 18.01° line in the 2D map) reveals a broad dispersion in NW tilts, in average estimated at ±2.85˚. (d) χ-φ pole figure, taken around AlN(0002) Bragg reflection, reveals a high isotropy in NWs' alignment on $SiO_2$.

A two-dimensional ω/2Θ – ω scan, performed in the proximity of the (111) Bragg reflection of the underlying Si(111) substrate (Fig. 4(a)), reveals a relatively intense and broad (0002) Bragg reflection of the AlN epilayer. This result proves that a significant fraction of NWs grows vertically aligned to the Si(111) crystallographic direction (AlN(0002) ∥ Si(111)), i.e. perpendicular to the $SiO_2$ substrate surface, as observed by RHEED (Fig. 1(a)). No Bragg reflections related to $SiO_2$ interlayer have been observed, confirming amorphous nature of this interlayer (in agreement with TEM).

To get further insight into the grown AlN epilayer, two one-dimensional scans have been performed. First, a θ/2θ scan featured in Fig. 4(b) (this scan corresponds to the "horizontal" ω = 0° line in the 2D map featured in Fig. 4(a)) reveals the angular position of the AlN(0002) peak at $\theta_{0002} \approx 18.01°$. By applying formula for Bragg diffraction, the distance between the corresponding $d_{0002}$ planes ($d_{0002} = \lambda_X / 2 \sin \theta_{0002}$) and the AlN c-lattice constant (c = $2d_{0002}$) are found: c = 4.98 Å. The found value of the c lattice constant is widely reported as the value corresponding to the fully relaxed AlN, confirming thus that AlN NWs grow virtually free of strain, as expected for self-assembled columnar nanostructures [16]. Second, an ω-rocking scan featured in Fig. 4(c) (this scan corresponds to the "vertical" ω/2Θ = 18.01° line in the 2D map featured in Fig. 4(a)) confirms a high dispersion in NWs tilts; defining an average NWs' tilt simply as one half of the ω-rocking full width at half maximum (Fig. 4(c)), we estimate it at ±2.85˚ with respect to the direction perpendicular to the $SiO_2$. Note that the high value of NWs' tilts found by ω-rocking scans is in a qualitative agreement with the annular shape of Bragg spots observed *in-situ* by RHEED (Fig. 1(a)). Finally, χ-φ pole figure around AlN(0002) Bragg reflection has been recorded (Fig. 4(d)) [16]. This measurement confirms that the NWs tilt is very uniformly distributed and further evidences a lack of any crystallographic preference in in-plane crystallographic directions. Note that the high isotropy of NWs orientation found by χ-φ two-dimensional scans agrees with the RHEED profile which is found invariant to the sample rotation (Fig. 1(a)).

It is important to note that the random NW twist introduces randomness not only in the position of "vertical" AlN crystalline planes (($10\bar{1}0$) and ($11\bar{2}0$)), but also in the position of the "inclined" ones (($10\bar{1}n$) and ($11\bar{2}n$)). Thus, XRD measurements were successfully performed only on "horizontal" crystalline planes i.e. on those not affected by the random twist. In other words, only symmetric Bragg reflections, such as (0002), were resolved, whereas the intensity of the asymmetric Bragg reflections (($10\bar{1}n$) and ($11\bar{2}n$)) remained below the detection level. Consequently, a more detailed study of NWs strain could not have been performed by XRD [16].

To gain an insight into optical properties of the grown AlN NWs, photoluminescence (PL) measurements were performed with 193 nm line of excimer ArF laser. Figure 5(a) shows the PL spectra obtained in the 2 – 300 K temperature range. The low temperature measurement resolves AlN near band edge (NBE) emission (at 6.03 eV) and its intense 1st order phonon replica (separated 103 meV in the spectrum). Both peaks show significant red shift and thermal quenching with increasing temperature. In particular, the AlN NBE emission, shifts from 6.03 eV to 5.95 eV, whereas its full width at half maximum (FWHM) increases from 20.4 to 48.4 meV. Details about the NBE emission evolution with increasing temperature are summarized in fig. 5(b).

To precisely determine NBE emission spectral position and its FWHM, the results shown in Fig. 5(a) have been fitted to Gaussian functions. The temperature dependent change in conduction-to-valence band transition energy is commonly fitted with either Varshni or Bose–Einstein equations [17]. The Varshni equation states that: $E(T) = E(0) - \alpha T^2/(\beta + T)$, where $E(T)$ and $E(0)$ are energies of the analyzed transitions, at the measured temperature T and at 0 K, and α and β are the Varshni coefficients. The theoretical basis for the Varshni equation, however, is unclear. Consequently, an alternative Bose–Einstein equation, which takes into consideration the renormalization of electronic states due to their interaction with phonons with average phonon frequency, is often used: $E(T) = E(0) - 2a_B/(e^{\theta/T} - 1)$, where $a_B$ is the strength of the average exciton–phonon interaction and θ is the average phonon frequency in the unit of temperature [17]. Both Varshni and Bose-Einstein eq. fits are shown in Fig. 5b. Similar to results of other authors, we find that Bose-Einstein eq. fits the experimental data significantly better (adjusted $R^2$ = 0.9995) than Varshni eq. (adjusted $R^2$ = 0.9916, Fig. 5b).

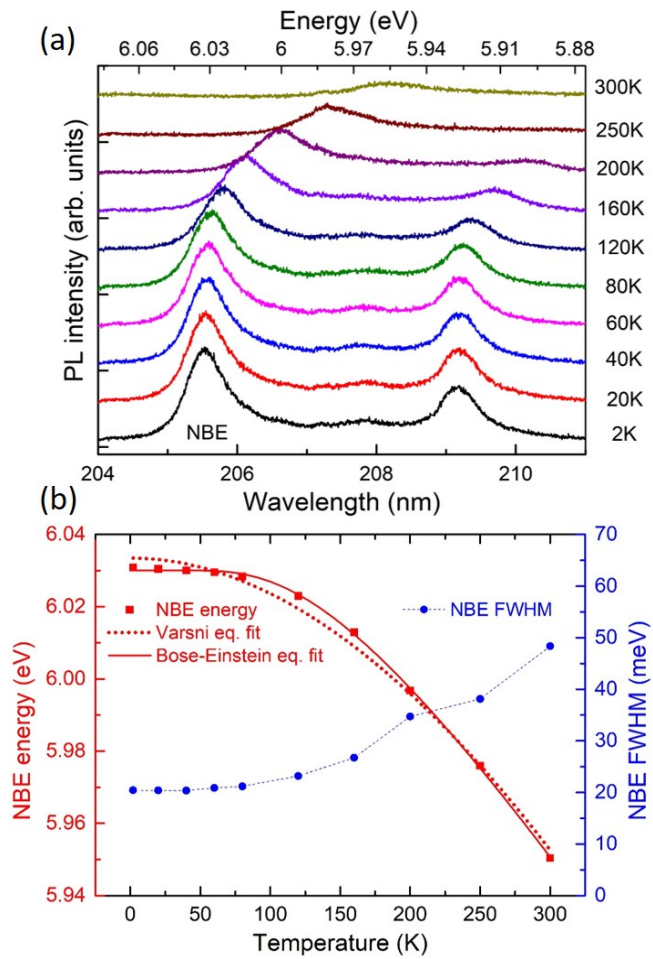

Figure 5. (a) Temperature dependent PL measurements of AlN nanocolumns reveal NBE emission together with its intense 1st order phonon replica. Both resolved peaks show common red shift and quenching with increasing temperature. (b) Red shift and quenching of the NBE emission with increasing temperature.

Table I summarizes the relevant aspects of photoluminescence of AlN layers grown in the different morphologies (compact vs. columnar epilayers) and using different growth techniques: physical vapor transport (PVT) vs. metal organic vapor phase epitaxy (MOVPE) / metal organic chemical vapor deposition (MOCVD) vs. molecular beam epitaxy (MBE). In particular, this table addresses: (i) NBE excitonic features at low temperature (< 10 K) and (ii) NBE PL evolution with increasing temperature (up to 300 K) [12, 18 - 23].

Concerning (i), it is interesting to note that excitonic features of the NBE emission have been resolved on the compact [18 - 22], but not on the columnar AlN epilayers [12, 23]. Concerning the compact AlN layers themselves, when fabricated homoepitaxially making use of physical vapor transport (PVT) technique, very sharp NBE excitonic lines have been reported (best values ≈1 meV) [18, 19]; when, on the other hand, AlN was grown heteroepitaxially on foreign substrates by metal organic vapor phase epitaxy (MOVPE) / metal organic chemical vapor deposition (MOCVD) techniques, then typically two peaks, attributed to free and a bound exciton, with significant FWHM (> 5 meV) have been reported [20 - 22]. This result is not surprising since heteroepitaxy implies the formation of extended defects that propagate into AlN epilayer, affecting its optical properties. The broadening is even stronger in the case of columnar AlN, thus PL measurements on these structures typically report one single NBE emission band, in which free and bound excitons are probably mixed together [12, 13]. The FWHM of this band is found at ≈20 meV or higher.

**Table I**. Comparison of NBE emission properties (as determined by PL measurements) of AlN layers grown with different morphologies and by different growth techniques [12, 18 - 23].

| Layer properties | | NBE emission properties as determined by PL | | | | Emission line traced in TDPL / / Reference |
|---|---|---|---|---|---|---|
| Epilayer type (growth technique) / substrate (growth technique) | | Low temperature PL (≤ 10 K) | Temperature dependent PL: Bose-Einstein eq. fitting parameters | | | |
| | | NBE features / FWHM | $E(0)$ [$eV$] | $a_B$ [$meV$] | $\theta$ [$K$] | |
| Bulk/2D layers homo-epitaxial | Bulk AlN (PVT) / AlN (PVT) | $FX_A$ and four $D^OX$s resolved / narrowest $D^OX$: **1.4 meV** | 6.0343 | 200 | 558 | $FX_A$ traced / M. Feneberg et al (2010) |
| | 2D AlN (MOVPE) / AlN (PVT) | $FX_A$ and two $D^OX$s resolved / narrowest $D^OX$: **1 meV** | 6.0411 | 170 | 510 | $FX_A$ traced / M. Funato et al (2012) |
| 2D layers hetero-epitaxial | 2D AlN (MOCVD) / sapphire | Two peaks (higher energy peak at 6.033 eV) / **11.4 meV** | | | | J. Li et al (2002) |
| | 2D AlN (MOCVD) / sapphire | Single peak identified as $FX_A$ / **16 meV** | 6.023 | 670 | 1000 | $FX_A$ traced / K.B. Nam et al (2004) |
| | 2D AlN (MOCVD) / 6H-SiC | Two peaks identified as $FX_A$ and $D^OX$ / ≈ **5 meV** | 5.985 | 72 | 345 | $D^OX$ traced / G.i M. Prinz et al (2007) |
| NWs hetero-epitaxial | Columnar AlN (MBE) / $SiO_2$ | Single emission band at 6.02 eV / **> 40 meV** | - | - | - | O. Landré et al (2010) |
| | Columnar AlN on GaN NWs (MBE) / Si(111) | Single emission band at 6.03 eV / **21 meV** | - | - | - | Q. Wang et al (2014) |
| | Columnar AlN / $SiO_2$ | Single emission band / **20.4 meV** | 6.0301 | 137 | 249 | Emission band position traced / This work |

It is important to note that NBE emission of ensembles of SA AlN NWs is significantly different from NBE emission of ensembles of GaN NWs; in the latter case the NBE emission is typically composed from several very narrow excitonic lines (best values typically below 1 meV) [24, 25].

The MBE growth of SA AlN NWs differs in three major aspects from the growth of their SA GaN NW counterparts, some of them being speculated as a possible origin for experimentally observed strong broadening of excitonic lines [12]. First, (i) AlN NWs are grown on $SiO_2$, whereas GaN NWs are typically grown on Si substrate. It is known that $SiO_2$ can thermally dissociate at temperatures used for AlN growth (920 – 940 °C), thus oxygen impurities can be easily incorporated into AlN. Second, (ii) AlN NWs are grown without any epitaxial relationship to the underlying Si(111) substrate whereas GaN NWs do preserve epitaxial relationship to the underlying Si(111) substrate [26, 27], their coalescence leading to the formation of irregular (non-hexagonal) cross-section NWs (as observed by SEM). Thus, a dense network of extended defects is expected at AlN NWs coalescence joints. Finally, (iii) unlike GaN, AlN is prone to oxidation, consequently AlN NWs are wrapped by a thin layer of amorphous oxide surface (as observed by STEM). However, at this stage, further study is necessary to determine the dominant mechanism for the observed broadening.

Concerning (ii) (NBE PL evolution with increasing temperature, up to 300 K), it is worth noticing that all reports, presented in table I, show very good fitting of the temperature dependent change in conduction-to-valence band transition energy to the Bose–Einstein equation. However, a significant dispersion in the reported fitting parameters $a_B$ and $\theta$ is observed. Here, we identify two possible reasons for such a significant dispersion. First, the Bose–Einstein fitting equation is derived within the frame of the renormalization of electronic states due to their interaction with phonons in the material [17]. A strong phonon replica in the PL spectrum is observed by many authors, however both the energy of the NBE – 1st order phonon replica separation as well as the physical origin of the dominant phonon modes remain a matter of debate. Concerning only AlN nanowires grown by MBE, the reported energy separation varies from 100 to 110 meV [12, 23]; in particular, Landré *et al* report the phonon replica at energy of ≈110 mV and attribute it to LO phonon modes whereas Wang *et al* report the phonon replica at energy of ≈100 meV and attribute it to SO phonon modes. Hereby, we report this value to be 103 meV, i.e. a value in between the two previous ones. It is also worth noticing that Wang *et al* suggest that AlN morphology and Al surface oxidation could have strong influence on (and possibly change) the dominant phonon mode. The same hypothesis (that the dominant phonon mode is not necessarily the same in all reported AlN crystals) could also explain a significant dispersion in reported $a_B$ and $\theta$ fitting parameters. Second, because of its exponential form, the fitting formula ($E(T) = E(0) - 2a_B/(e^{\theta/T} - 1)$), is very sensitive to a slightest change of the fitting parameters (in particular, to a change of the parameter $\theta$). This makes the fitting procedure highly sensitive to a smallest error in input experimental data, that is, only a slight deviation in the experimental data set, leads to a significantly different $\theta$, and, consequently, $a_B$ parameter. This sole mathematical property of the fitting function could also explain the high dispersion of the reported fitting parameters.

In summary, SA AlN nanowires (NWs) are grown by PAMBE on SiO$_2$/Si (111) substrates. Using a combination of *in situ* RHEED measurements and properly selected one-dimensional and two-dimensional *ex situ* XRD measurements, we show that NWs grow preferentially perpendicular to the amorphous SiO$_2$ interlayer, but without any epitaxial relationship to the underlying Si(111) substrate, as expected. ω-rocking scans estimate NWs' tilt at ±2.85°, whereas two-dimensional χ-φ scans show randomness and uniformity of NWs' twist. Symmetric ω/2θ scans reveal AlN (0002) Bragg reflection at 18.01°, corresponding to c lattice constant of fully relaxed AlN (c = 4.98 Å). SEM reveals that the NWs diameter progressively increases as the growth proceeds, mainly due to the significant NWs coalescence, resulting finally in formation of columnar structures with non-hexagonal cross-section. STEM estimates their initial diameters in the 20 – 30 nm range. In addition, STEM evidences the formation of a thin (≈30 nm) polycrystalline AlN layer at the interface. As long as structural quality of AlN NWs is concerned, STEM reveals formation of extended columnar regions which grow with a virtually perfect metal-polarity wurtzite arrangement and only sporadic appearance of extended crystalline defects, such as stacking faults. The combination of STEM and EELS, reveals that NWs are wrapped by a thin (1 – 2- nm) aluminum oxide layer. Low temperature photoluminescence measurements reveal NBE emission at 6.03 eV (at 2 K) and its intense phonon replica separated 103 meV in the spectrum. The significant FWHM of the NBE emission, found at ≈20 meV (at 2 K), suggests that free and bound excitons are mixed together within this single emission band. The temperature dependent change in conduction-to-valence band transition energy (determined by PL measurements in the 2K to 300 K range), is very well fitted to the Bose–Einstein equation. The fitting parameters are found at $a_B = 137\ meV$ and $\theta = 249\ K$. Finally, we show an insightful comparison of NBE emission, as determined by PL measurements, of AlN layers grown with different morphologies (bulk, thin layer or columnar) and by different experimental techniques (PVT, MOCVD/MOVPE and MBE).

Acknowledgements. We acknowledge Dr. A. Peral and Dr. C. del Cañizo from Instituto de Energía Solar, Universidad Politécnica de Madrid, for thermal deposition of SiO$_2$ and thank Dr. N. Vukmirović from Institute of Physics, Belgrade, for fruitful discussions. Electron microscopy observations were carried out at the National Center for Electron Microscopy (ICTS ELECMI) at Universidad Complutense de Madrid. This work has been financially supported by Spain's Ministry of Science and Innovation under Grants No. RTI2018-097895-B-C43 and RTI2018-097338-B-100.**References**